\documentclass[aps,prb,twocolumn,showpacs,showkeys,superscriptaddress]{revtex4}
\usepackage{epsfig}
\usepackage{hyperref}

\begin{document}

\def\il{I_{low}} 
\def\iu{I_{up}} 
\def\eeq{\end{equation}}
\def\ie{i.e.}  
\def\etal{{\it et al. }}  
\def\prb{Phys. Rev. {B }}
\def\pra{Phys. Rev. {A }} 
\def\prl{Phys. Rev. Lett. }
\def\pla{Phys. Lett. } 
\def\pb{Physica B}
\def\ajp{Am. J. Phys. }
\def\apl{Appl. Phys. Lett. } 
\def\jpc{J. Phys. C } 
\def\rmp{Rev. of Mod. Phys. } 
\def\jap{J. Appl. Phys. } 
\def\mpl{Mod. Phys. Lett. {B }} 
\def\ijmp{Int. J. Mod. Phys. {B }} 
\def\ijp{Ind. J. Phys. }
\def\ijpap{Ind. J. Pure Appl. Phys. }
\def\ibmjrd{IBM J. Res. Dev. }
\def\pjp{Pramana J. Phys. }

\title{Survival of $\Phi_{0}/2$ periodicity in presence of incoherence
  in asymmetric Aharonov-Bohm rings} 
\author{Colin Benjamin}
\email{colin@iopb.res.in} 
\affiliation{Institute of Physics,
  Sachivalaya Marg, Bhubaneswar 751 005, Orissa, India}
\author{Swarnali Bandopadhyay} 
\email{swarnali@bose.res.in}
\affiliation{S N Bose National Center for Basic Sciences, JD Block,
  Sector III, Salt Lake City, Kolkata 700098, India} 
\author{A. M. Jayannavar}
\email{jayan@iopb.res.in} 
\affiliation{Institute of Physics,
  Sachivalaya Marg, Bhubaneswar 751 005, Orissa, India}

\date{\today}

\begin{abstract}
  Magneto conductance oscillations periodic in flux with periodicity
  $\Phi_{0}$ and $\Phi_{0}/2$ are seen in asymmetric Aharonov-Bohm
  rings as a function of density of electrons or Fermi wave vector.
  Dephasing of these oscillations is incorporated using a simple
  approach of wave attenuation. In this work we study how the
  excitation of the $\Phi_{0}/2$ oscillations and the accompanying
  phase change of $\pi$ are affected by dephasing. Our results show
  that the $\Phi_{0}/2$ oscillations survive incoherence, i.e.,
  dephasing, albeit with reduced visibility while incoherence is also
  unable to obliterate the phase change of $\pi$.
 
\end{abstract} 

\pacs{72.10.-d, 73.23.-b, 05.60.Gg, 85.35.Ds}

\keywords{D. Electron Transport, A. Nanostructures, D. Aharonov-Bohm
  oscillations, D. Dephasing} 

\maketitle

The $\Phi_{0}/2$ periodicity was a puzzle in mesoscopic physics in its
early days. Among the first experiments\cite {sharvin} which were
purported to measure the magneto resistance oscillations in normal
metal cylinders, observed a $\Phi_{0}/2$ periodicity. However,
theoretical calculations\cite{th,butipra,gefenprl} on strictly
one-dimensional normal metal ballistic rings argued that only
$\Phi_{0}$ periodicity should be observed. The experiment which
observed these $\Phi_{0}/2$ oscillations were backed by theoretical
work which predicted these based on weak localization\cite{AAS,wash}.
In the recent works of Pedersen, et.al.,\cite{peders} and Hansen,
et.al.,\cite{hansen}, the AB effect in a one dimensional
$GaAs/Ga_{0.7}Al_{0.3}As$ ring at low magnetic fields has been
investigated. In their work they observe the fundamental $\Phi_0$
periodicity in the magneto-conductance as expected. Moreover, as the
density (in effect the Fermi energy) is varied they observe phase
shifts of $\pi$ in the magneto conductance oscillations and $\Phi_0/2$
periodicity at particular values of the Fermi energy. They have found
good agreement of their results with the completely phase coherent
transport theory\cite{squid} of electrons in an asymmetric
Aharonov-Bohm ring in the single channel regime. Asymmetry of the AB
ring was crucial in understanding these observations. Such behavior
has also been observed in an earlier experiment\cite{yacoby}, and has
generated a lot of interest in relation to the problem of phase
measurement.

The endeavor of this work is not on the origin of the $\Phi_{0}/2$
periodicity but on the effect of inelastic or phase breaking
scattering on these.  Our results indicate that the phase shift of
$\pi$ in AB oscillations and halving of the fundamental $h/e$
periodicity survives in-spite of dephasing albeit with reduced
visibility in AB oscillations. There are many ways to
phenomenologically model inelastic scattering in mesoscopic devices.
Among the first was by B\"{u}ttiker \cite{butinelas}who considered an
electron reservoir coupled by a lead to a mesoscopic system as a phase
breaker or inelastic scatterer (voltage probe). This approach has been
widely used to investigate the effect of dephasing on the conductance.
This method which uses voltage probes as dephaser's is interesting
because of it's conceptual clarity and it's close relation to
experiments. It provides a useful trick to simulate lack of full
coherence in transport properties. This method of addressing the
problem of dephasing has the advantage that inelastic phase
randomizing processes can be incorporated by solving an elastic time
independent scattering problem. Beyond B\"{u}ttiker's model, optical
potential,\cite{ferry,jayan} and wave attenuation (stochastic
absorption) models \cite{colin,joshi} have also been used to simulate
dephasing. However in the aforesaid models energy relaxation and
thermal effects\cite{mortensen} are ignored. Thermal effects can be
incorporated by taking into account thermal distribution (Fermi-Dirac
function) of electrons.  In mesoscopic systems, transmission functions
are more often than not constant over the energy range wherein
transport occurs (at low temperatures) and one can ignore energy
relaxation or ``vertical flow''\cite{datta} of electron carrier's in
these systems. Brouwer and Beenakker have corrected some of the
problems associated with voltage probe and optical potential models,
(see Refs.[\onlinecite{brouwer,colin}] for details), and given a
general formalism for calculating the conductance(G) in the presence
of inelastic scattering. Furthermore, methods based on optical
potentials and wave attenuation can make use of this above formalism.
In this work we use the method of wave attenuation.
 
\begin{figure*}
\protect\centerline{\epsfxsize=5.6in\epsfbox{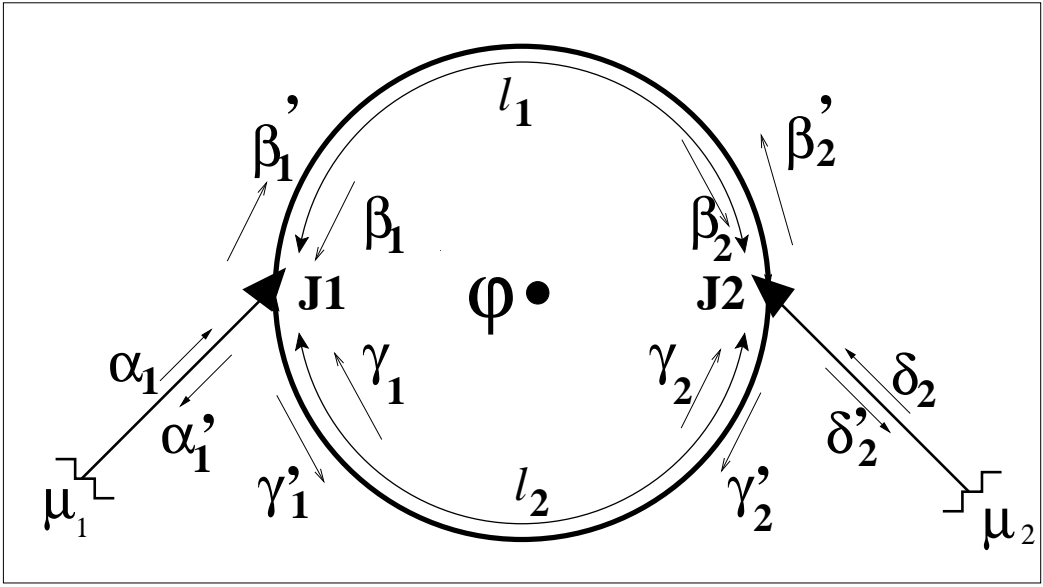}}
\caption{Aharonov - Bohm ring geometry.}
\end{figure*}

This method of wave attenuation has been used earlier to study
dephasing of AB oscillations\cite{colin} and calculating sojourn times
in quantum mechanics\cite{colinssc}. The wave attenuation model has
been shown to be better than the optical potential model (which has in
built spurious scattering)\cite{colin}. We use the well known S-Matrix
method to calculate the conductance and therein we see the
$\Phi_{0}/2$ periodicity as also the phase change of $\pi$ across such
an excitation of the $h/2e$ oscillations.  The system we consider, is
shown in FIG.~1, is an asymmetric Aharonov-Bohm loop with upper and
lower arm lengths $l_1$ and $l_2$ and circumference $L=l_{1}+l_{2}$,
coupled to two leads which in turn are connected to two reservoirs at
chemical potentials $\mu_1$ and $\mu_2$. Inelastic scattering is
assumed to be absent in the leads while it is present in the
reservoirs, and in the loop we introduce incoherence via wave
attenuation to simulate inelastic scattering.  The S matrix for the
left coupler yields the amplitudes
$O_{1}=(\alpha_{1}^\prime,\beta_{1}^\prime,\gamma_{1}^\prime)$
emanating from the coupler in terms of the incident waves
$I_1=(\alpha_{1},\beta_{1},\gamma_{1})$, and for the right coupler
yields the amplitudes
$O_{2}=(\delta_{2}^\prime,\beta_{2}^\prime,\gamma_{2}^\prime)$
emanating from the coupler in terms of the incident waves
$I_2=(\delta_{2},\beta_{2},\gamma_{2})$. The S-matrix for either of
the couplers\cite{butipra} is given by-

\[S=\left(\begin{array}{ccc}
-(a+b)       & \sqrt\epsilon&\sqrt\epsilon\\
\sqrt\epsilon& a            &b            \\
\sqrt\epsilon& b            &a            
\end{array} \right) \]

with $a=\frac{1}{2}(\sqrt{(1-2\epsilon)} -1)$ and
$b=\frac{1}{2}(\sqrt{(1-2\epsilon)} +1)$. Herein, $\epsilon$ plays the
role of a coupling parameter. The maximum coupling between reservoir
and loop is $\epsilon=\frac{1}{2}$, and for $\epsilon=0$, the coupler
completely disconnects the loop from the reservoir. Inelastic
scattering in the arms of the AB interferometer is taken into account
by introducing an attenuation constant per unit length in the two arms
of the ring, i.e., the factors $e^{-\alpha l_1}$ (or $e^{-\alpha
  l_2}$) in the free propagator amplitudes, every time the
electron\cite{colin,datta} traverses the upper (or lower) arms of the
loop (see Fig.~1).

The waves incident into the branches of the loop are related by the S
Matrices \cite{cahay}for upper branch by-

\[\left(\begin{array}{c}
\beta_1\\
\beta_2\\
\end{array} \right) \ =\left(\begin{array}{cc}
0     & e^{ikl_1} e^{-\alpha l_1} e^\frac{-i \theta l_1}{L}\\
e^{ikl_1} e^{-\alpha l_1} e^\frac{i \theta l_1}{L} & 0 \\
\end{array} \right) \left(\begin{array}{c}
\beta_1^\prime\\
\beta_2^\prime
\end{array} \right)\]  
and  for lower branch-

\[\left(\begin{array}{c}
\gamma_1\\
\gamma_2\\
\end{array} \right) \ =\left(\begin{array}{cc}
0     & e^{ikl_2} e^{-\alpha l_2} e^\frac{i \theta l_2}{L}\\
e^{ikl_2} e^{-\alpha l_2} e^\frac{-i \theta l_2}{L} & 0 \\
\end{array} \right) \left(\begin{array}{c}
\gamma_1^\prime\\
\gamma_2^\prime
\end{array} \right)\]

These S matrices of course are not unitary $
S(\alpha)S(\alpha)^\dagger\neq 1$ but they obey the duality relation $
S(\alpha)S(-\alpha)^\dagger= 1$. Here $kl_1$ and $kl_2$ are the phase
increments of the wave function in absence of flux.  $\frac{\theta
  l_1}{L}$ and $\frac{\theta l_2}{L}$ are the phase shifts due to flux
in the upper and lower branches.  Clearly, $\frac{\theta
  l_1}{L}+\frac{\theta l_2}{L}=\frac{2\pi\Phi}{\Phi_0} $, where $\Phi$
is the flux piercing the loop and $\Phi_0$ is the flux
quantum$\frac{hc}{e}$. The transmission and reflection coefficients
are given as follows-
$T_{21}=|\frac{\delta_{2}^\prime}{\alpha_{1}}|^2$,
$R_{11}=|\frac{\alpha_{1}^\prime}{\alpha_{1}}|^2$,
$R_{22}=|\frac{\delta_{2}^\prime}{\delta_{2}}|^2$,
$T_{12}=|\frac{\alpha_{1}^\prime}{\delta_{2}}|^2$ wherein wave
amplitudes $\delta_{2}^\prime,\delta_{2},\alpha_{1}^\prime,\alpha_{1}$
are as depicted in FIG.~1.

The transmission coefficient $T_{21}$ from reservoir 1 to 2 is not
symmetric under flux reversal which is in contradiction with Onsager's
symmetry condition, and is due to the fact that current conservation
as also unitarity have been violated (due to wave attenuation). As, is
well known there can be real absorption of photons but there cannot be
any real absorption of electrons. The absorption is interpreted as
electron scattering into different energy channels and the way these
electrons are re-injected back into the system becomes
important\cite{butiprama,pareek}.  Following the earlier treatments
(see the details in Refs. [\onlinecite{brouwer,colin}]), the
conductance in dimensionless form after proper re-injection of
carriers is given by -
\begin{eqnarray}
G=T_{21}+\frac{(1-R_{11}-T_{21})(1-R_{22}-T_{21})}{1-R_{11}-T_{21}+1-R_{22}-T_{12}}.
\end{eqnarray}

The first term in Eq.~1, i.e., $T_{21}$ represents the phase coherent
contribution, while the second term accounts for electrons that are
re-injected after inelastic scattering.  This represents the phase
incoherent contribution to the conductance. $G$ respects Onsager's
symmetry $G(\Phi)=G(-\Phi)$, and thus the phase of AB oscillations can
only change\cite{yacoby} by $\pm\pi$.

\begin{figure*} 
\protect\centerline{\epsfxsize=7.5in\epsfbox{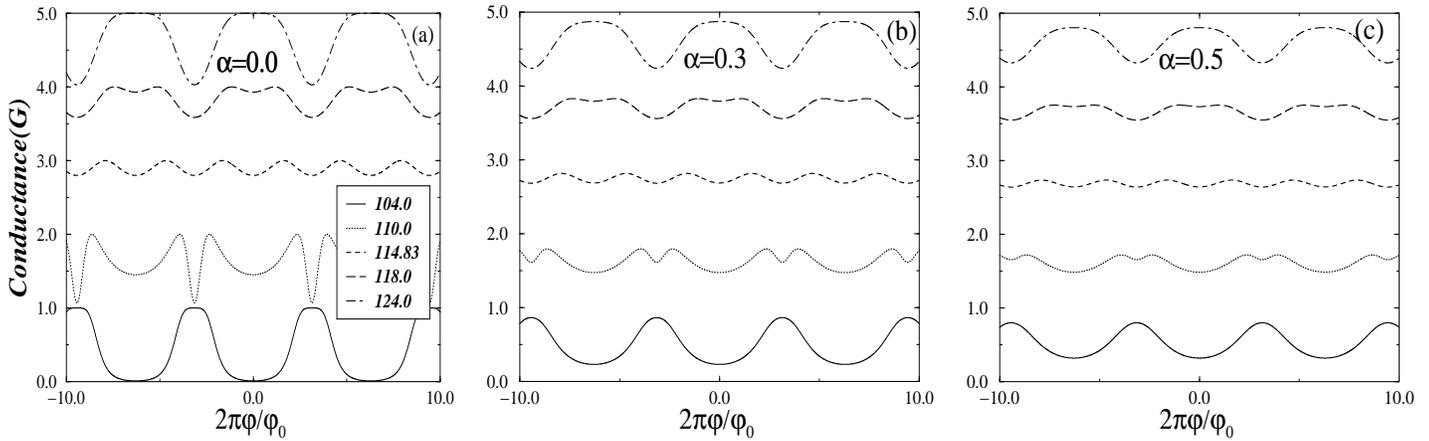}}
\caption{Conductance ($G$) for lengths $l_{1}/L=0.425$, $l_{2}/L=0.575$
  and coupling parameter $\epsilon=0.5$ (strong coupling) for
  different values of the Fermi wave-vector $k_{f}L$. The legend in
  FIG.~2(a) remains same for 2(b) and 2(c).}
\end{figure*}

\begin{figure*} 
\protect\centerline{\epsfxsize=7.5in\epsfbox{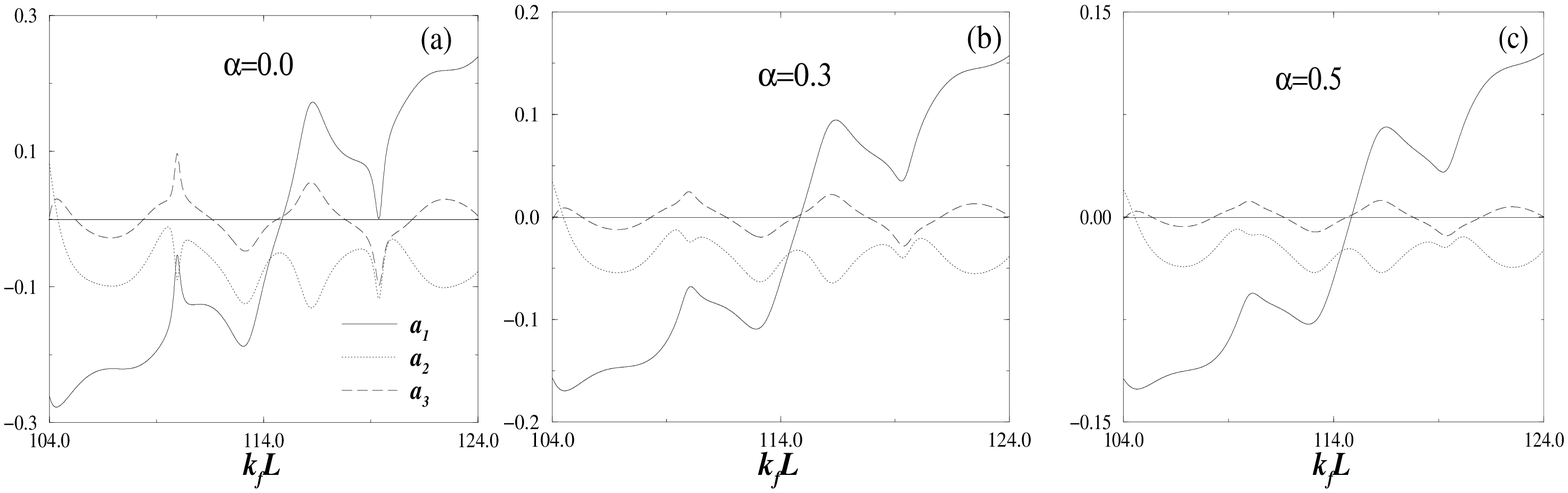}}
\caption{Harmonics for lengths $l_{1}/L=0.425$, $l_{2}/L=0.575$
  and coupling parameter $\epsilon=0.5$ (strong coupling) as a
  function of the dimensionless Fermi wave-vector $k_{f}L$. The legend
  in FIG.~3(a) remains same for 3(b) and 3(c).}
\end{figure*}

As previously mentioned our interest in this work is to observe the
effect of incoherence on the $\Phi_{0}/2$ oscillations in single
channel ballistic rings. We choose an asymmetric AB ring with degree
of asymmetry denoted by the difference in arm lengths
$\Delta=l_1-l_2=0.15$, and circumference $L=1.0$ in accordance with
the experimental realization as in Ref. [\onlinecite{peders}]. The
change in Fermi energy of injected electrons implies varying the
density of electrons in the system. So, when we scan the whole range
of the dimensionless wave vector $k_fL$ from $0.0$ to $200.0 $ we come
across $\Phi_{0}/2$ periodicities at particular values of the Fermi
wave vector $k_fL$, notably at $10.83, 114.8302$ and $136.5$. We now
restrict ourselves to the particular range(Fermi energy) and
parameters (length's and coupling) corresponding to the experimental
situation studied earlier as in Refs. [\onlinecite{peders,hansen}]. In
our treatment $\alpha$ represents the incoherence parameter (degree of
dephasing). The plot of the dimensionless conductance $G$ as a
function of flux in the range $104.0 \le k_{f}L \le 124.0$, with
degree of incoherence $\alpha=0$ is shown in FIG.~2(a). Similarly in
FIG.~2(b) and 2(c) we plot $G$ for $\alpha=0.3$ and $\alpha=0.5.$, for
the same system parameters and range of $k_{f}L$.  The plots for
$k_{f}L > 104.0$ are each shifted by a factor of 1 for clarity. The
$\Phi_{0}/2$ periodicities are clearly marked at $k_{f}L=114.8302$,
and also across this range of $k_{f}L$ and excitation of the $h/2e$
harmonic, phase changes by $\pi$. Thus phase shift of $\pi$ along with
halving of the fundamental $h/e$ period is clearly seen as a function
of Fermi-wavevector (density) consistent with the experimental
observations. Importantly, this observed behavior survives dephasing
with reduced visibility, therefore the observed results need not be
attributed to complete phase coherence in the system.  One conclusion
which can be drawn from the afore drawn figures is that incoherence
reduces the visibility of AB oscillations as expected. However, this
dephasing is unable to shift the position of the $\Phi_0/2$
oscillations noticeably, for the chosen coupling parameter.

The reason why we observe $\Phi_{0}/2$ periodic oscillations at these
particular values of $k_{f}L$ is because at these values both $h/e$ as
well as $h/3e$ harmonics are extremely weak as also the higher
harmonics and therefore exclusive $\Phi_{0}/2$ oscillations are seen.
The $k_{f}L$ values wherein exclusive $\Phi_{0}/2$ oscillations are
seen are at $k_{f}L=10.8335, 114.8302$ and $136.5$, in the range $0.0<
k_{f}L< 200.0$ for the same physical parameters. In FIG.3(a),(b) and
(c), we plot the harmonics as a function of the dimensionless Fermi
wave-vector $k_{f}L$ for $\alpha=0.0,0.3$ and $0.5$. The harmonics are
calculated as follows-

\begin{eqnarray}
a_{n}=\frac{1}{\pi}\int_{0}^{2\pi} G cos(n\theta) d\theta
\end{eqnarray}

At the '$k_{f}L$' value, wherein $\Phi_{0}/2$ oscillations dominate,
the first and third harmonic's do not contribute at all to the
conductance as can be seen from the FIG's.3(a)-(c). We observe that
increasing dephasing ($\alpha$) does not noticeably shift the
'$k_{f}L$' value, wherein $\Phi_{0}/2$ oscillations dominate. We also
see that the higher harmonic $a_{3}=h/3e$ goes faster to zero and
therefore these contributions are washed out and $\Phi_{0}/2$
oscillations survive dephasing albeit with reduced strengths. The fact
that the Fermi-wavevector $k_{f}L$ (at which $\Phi_0/2$ oscillations
occur) does not noticeably shift is peculiar to the coupling parameter
chosen, which for the above cases is $0.5$(maximal coupling). However,
for some other physical parameters there may be a small shift in
Fermi-wavevector $k_{f}L$ with increasing incoherence. For example,
for the case $\epsilon=0.44$ (waveguide coupling) the $\Phi_{0}/2$
oscillations are observed at $k_{f}L=52.0$ at $\alpha=0.0$, for the
same length parameters as in FIG.~2, but when this incoherence
parameter is increased we see these oscillations are shifted to
different values of $k_{f}L$, e.g., for $\alpha=0.5$ these are seen at
$k_{f}L=51.95$. For this coupling too we indeed observe phase change
of $\pi$ in AB oscillations along with period halving, consistent with
our previous observations. Shifts in Fermi-wavevector are small for
maximal coupling but when coupling strength is decreased these shifts
become more noticeable.

\vskip 0.0in 

In conclusion, we have observed $\Phi_{0}/2$ oscillations as we vary
the density of electrons which is similar to varying the Fermi wave
vector consistent with experimental observations. The $\Phi_{0}/2$
oscillations are shifted by dephasing (noticeably small for maximal
coupling), apart from the reduction of their strengths.  The phase
change of $\pi$ which occurs across the excitation of $h/2e$
oscillations is seen to be independent of dephasing. Thus complete
phase coherence of electron over the entire sample is not necessary to
observe these effects.

\acknowledgments
One of us SB thanks the Institute of Physics, Bhubaneswar for hospitality.

\end{document}